\documentclass[twocolumn,showpacs,preprintnumbers,amsmath,pre]{revtex4}

\usepackage{graphicx}
\usepackage{dcolumn}
\usepackage{bm}
\usepackage{amssymb}

\begin{document}

\title{Thermodynamic limit of the 
first-order phase transition in the Kuramoto model}
\author{Diego Paz\'o}
\email{pazo@pks.mpg.de}
\affiliation{Max-Planck-Institut f{\"u}r Physik komplexer Systeme, 
N{\"o}thnitzer Stra{\ss}e 38, 01187 Dresden, Germany}

\date{\today}

\begin{abstract}

In the Kuramoto model, a uniform distribution of the natural
frequencies leads to a first-order (i.e., discontinuous)
phase transition from incoherence to synchronization, 
at the critical coupling parameter $K_c$.
We obtain the asymptotic dependence 
of the order parameter above criticality: 
$r-r_c \propto (K-K_c)^{2/3}$.
For a finite population, we demonstrate
that the population size $N$ may be included into
a self-consistency
equation relating $r$ and $K$
in the synchronized state. 
{We analyze the convergence to the thermodynamic limit 
of two alternative schemes to set the natural frequencies.
Other frequency distributions different
from the uniform one are also considered.}

\end{abstract}
\pacs{05.45.Xt}
\maketitle
\section{Introduction}

Synchronization is a universal phenomenon
that plays 
an important role in all natural sciences as
well as in technology \cite{PRK,Blekhman}. 
In particular, the synchronization 
of populations of globally coupled 
oscillators with distributed natural frequencies 
has been an object of study since very early times,
mainly in a biological context~\cite{Winfree}.
Later, it has found application in other areas, 
such as Josephson junctions~\cite{wiesenfeld}, 
nanomechanics~\cite{alex}, etc. 
When increasing the coupling parameter,
these systems undergo transitions 
from a totally incoherent state to 
a partially coherent state where part of the population
becomes entrained sharing the same frequency. 
Interestingly, there
are several analogies with the phase 
transitions in statistical mechanics~\cite{daido90}.
Thus, one may define an order parameter, that usually
grows continuously from zero (the incoherent state) 
when the coupling parameter exceeds a threshold value, 
analogously to a second-order phase transition.
Nonetheless, in some situations~\cite{inertia,inertiad,bonilla92},
mutual entrainment occurs in an abrupt way (a 
first-order phase transition). After an infinitesimal 
variation of the coupling strength 
a macroscopic (i.e.~order $N$) part of the population becomes 
synchronized. One may speculate that 
first-order phase 
transitions may be of 
interest for practical applications, if one pursues  
a system exhibiting an abrupt off/on switch.

The most simple example 
of a first-order phase transition
is found in the Kuramoto model~\cite{Kuramoto} when the
natural frequencies are uniformly distributed. In this
case, it is known that all the population becomes 
synchronized in a single step~\cite{vanhemmen,inertiad}.
We obtain here the asymptotic dependence of the order parameter
after criticality, which exhibits a critical exponent $2/3$. 

Still, one of the open problems in the Kuramoto model 
is to fully understand the finite-$N$ behavior. As we show below, 
the simplicity of the uniform frequency distribution allows
to  cope with finite-size effects in an original way,
providing analytic and numerical results. Some of 
these results apply to other frequency
distributions with compact support.

This paper is organized as follows.
In Sec.~\ref{km}, we present the Kuramoto model,
and show some numerical results that motivated
this work. In Sec.~\ref{cont} an infinite population
is considered, finding 
the asymptotic dependence of the order parameter
after criticality. 
In Sec.~\ref{discr}, we study 
finite ensembles, finding: a) an $N$-dependent formula for
the order parameter, b) different convergence rates to the
thermodynamic limit for alternative sampling schemes of the natural frequencies.
Section \ref{other} is devoted to analyzing
frequency distributions different from the uniform one.
Finally in Sec.~\ref{concl} 
the main conclusions of this work are summarized. 

\section{The Kuramoto Model}
\label{km}
The Kuramoto model is probably the most studied model of
synchronization in a population of oscillators 
with all-to-all coupling~\cite{rmp_kuramoto}. 
The state of each oscillator
is described only by a phase variable (this stems from
the the fact that, at small coupling,
only the phase of a self-sustained oscillator is affected by the 
interaction). The phase $\theta_j$ of each oscillator 
satisfies the following ODE:
\begin{equation}
\dot \theta_j = \omega_j + \frac{K}{N} \sum_{l=1}^N \sin( \theta_l- \theta_j)
\label{kuramoto_model}
\end{equation}
where $\omega_j$ are the natural frequencies, and $K$ 
is the parameter controlling the coupling strength. 

To quantify the state of synchronization,
Kuramoto proposed to use a complex-valued quantity (so-called
order parameter to emphasize the relation with phase transitions):
\begin{equation}
r e^{i\psi}= \frac{1}{N} \sum_{l=1}^{N} e^{i\theta_l} .
\label{op}
\end{equation}
It allows us to set the governing equation (\ref{kuramoto_model})
in the form:
\begin{equation}
\dot \theta_j = \omega_j + K r \sin( \psi - \theta_j) .
\label{mean_field}
\end{equation}
If the natural frequencies are
distributed (i.e.~$\omega_j \neq \omega_{j'}$), synchronization only appears above 
some coupling threshold.
Here, we consider the case of evenly spaced natural frequencies:
\begin{equation}
\omega_j=-\gamma+\frac{\gamma}{N}(2j-1)
\label{unif}
\end{equation}
which, like all symmetric frequency distributions 
can be assumed centered at zero
(by going into a rotating frame if necessary).
Throughout this paper, the numerical integration of the Kuramoto model 
[Eq.~(\ref{kuramoto_model})]
is carried out by means of a fourth-order Runge-Kutta method with time
step $\Delta t=0.1$.

\begin{figure}
\includegraphics[width=3.1in]{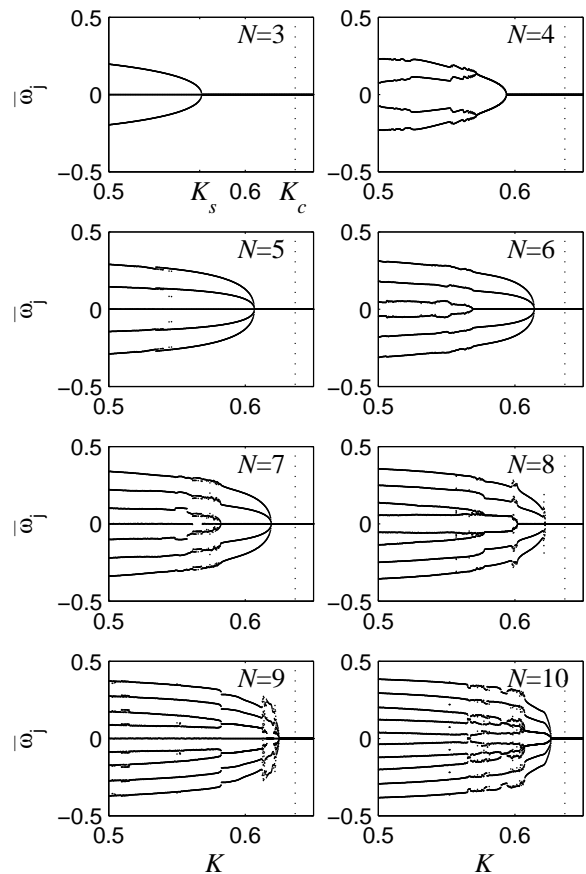}
\caption[]{Average frequencies ($\bar\omega_j$)
as a function of the coupling 
for different population sizes.
The natural frequencies are taken according
to (\ref{unif}) with $\gamma=1/2$. 
By $K_s$ we denote the value of $K$
for the first splitting bifurcation. The
critical point $K_c$ in the thermodynamic
limit ($N \rightarrow \infty$) is located at 
the dotted line.} \label{figN}
\end{figure}

Recently, Maistrenko {\em et al.}~\cite{maistrenko} 
studied the Kuramoto model with a small number of oscillators
and natural frequencies distributed uniformly (or close to that).
They found that the synchronized state
robustly splits into several clusters with 
different average frequencies.
It is shown there, and in Fig.~\ref{figN}, that for 
$N=3$ and $5$, the synchronized state splits directly into $N$ clusters.
But, for values of $N$ other
than 3 and 5 the scenario is not so simple.
As $N$ increases,
the number of splittings for going from one to $N$ clusters,
increases as well. Hereafter, we denote 
by $K_s$ the coupling at the frequency-splitting
from the synchronized state.
In congruence 
with the thermodynamic limit (see below), 
all the splittings [including the first one at $K=K_s(N)$] 
must accumulate at $K_c$ as $N\rightarrow \infty$.
$K_c$ is the 
abrupt transition point for an infinite population.

\section{Infinite population}
\label{cont}

In this section, we briefly study the 
Kuramoto model for a uniform frequency distribution.
Some of the formulas will be later compared
to those obtained for a finite population 
in Sec.~\ref{discr}. We also go one step further,
and deduce an explicit formula, with critical exponent 2/3, 
for the dependence of the order parameter after criticality.

In correspondence to the finite case, Eq.~(\ref{unif}), 
we consider a uniform density of the natural frequencies:
\begin{eqnarray}
g(\omega)=
\left\{
\begin{array}{cc}
\frac{1}{2\gamma} \mbox{ for } |\omega| \le \gamma \\
0 \mbox{ for } |\omega| > \gamma
\end{array}
\right. 
\label{g}
\end{eqnarray}

We first note that due to the invariance
under global rotation, for stationary solutions,
we can set a vanishing phase
for the order parameter in Eq. (\ref{op}), $\psi=0$,
without lack of generality. 
Kuramoto's classical analysis   
gives the order parameter equation:
\begin{equation}
r=\left< e^{i\theta} \right> = \left< \cos \theta  \right> 
\equiv \int_{-\infty}^{\infty} \cos \theta(\omega) \mbox{ } 
g(\omega)\mbox{ } d\omega  \mbox{ }.
\label{rcont}
\end{equation} 

In the totally locked regime, we obtain then:
\begin{eqnarray}
r&=& \int_{-\gamma}^\gamma g(\omega)
\sqrt{1-\frac{\omega^2}{K^2r^2}} d\omega  \mbox{ } \Rightarrow \label{rintegral} \\
r&=& \frac{1}{2} \sqrt{1-\frac{\gamma^2}{K^2r^2}}+\frac{Kr}{2\gamma}
\arcsin\left(\frac{\gamma}{Kr}\right) \mbox{ }.
\label{solvnum}
\end{eqnarray}

Eq.~(\ref{solvnum}) gives implicitely the dependence of $r$ on $K$.
A solution exists only for $Kr \ge \gamma$. 
At the critical point $K_cr_c=\gamma$,
the locked solution disappears with $r_c=\pi/4$. The corresponding value
of the  coupling is $K_c= 4\gamma/\pi$, that is precisely
the value where the incoherent solution $r=0$ becomes unstable
according to the classical result~\cite{strogatz91} for all unimodal 
distributions:
$K_c=2/(\pi g(0))$. Two remarks are in order. First, in contrast with
strictly unimodal distributions, the
transition is of first-order type: the order parameter ``jumps"
from zero to $r_c$. Second, at $K_c$ {\em all} the population becomes
entrained. This last
remark is quite important because it simplifies both numerical
and theoretical analyses. Also, note that when synchronized,
the oscillators' phases are spanned
along an interval (of length $\pi$ at $K=K_c$). When $K$
is increased $r$ grows from $r_c$ to $1$ in the 
$K\rightarrow \infty$ limit.

\begin{figure}
\includegraphics[width=3.1in]{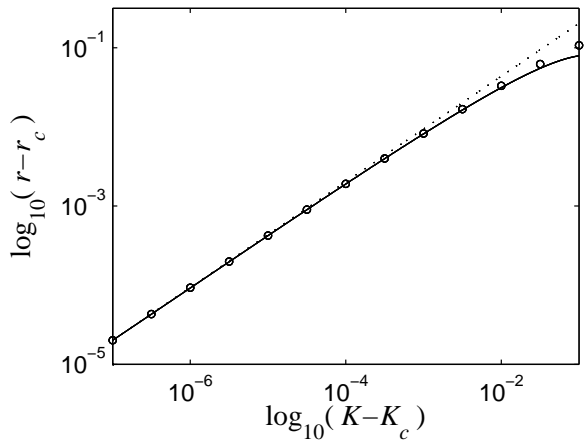}
\caption[]{Log-log dependence of $r$ on $K$ in a neighborhood
of criticality for $\gamma=1/2$. The circles correspond to
numerical solutions of the self-consistency condition (\ref{solvnum}), 
the solid line depicts Eq.~(\ref{2/3}), and the dotted straight line
arises taking the leading $\delta K^{2/3}$ term only.} \label{scaling}
\end{figure}

The first result of this paper is the dependence of $r$
on $K$, just above the phase transition. 
First of all, we make a change of variables onto Eq.~(\ref{rcont})
as usual (see e.g.~\cite{PRK,strogatz2000}):
\begin{equation}
r=Kr  
\int_{\theta_{min}}^{\theta_{max}} 
\cos^2\theta \mbox{ }
g(Kr\sin\theta) \mbox{ } d\theta .
\end{equation}
where $\theta_{max}$ ($\theta_{min}$) is the phase of the oscillator
with frequency $\gamma$ ($-\gamma$).
After an 
expansion above criticality:  
\begin{eqnarray}
K&=&K_c + \delta K \\
r&=&r_c + \delta r \\
\theta_{max} &=& -\theta_{min}=\pi/2 - \delta \theta .
\end{eqnarray}
and discarding the trivial incoherent solution $r=0$,
we get:
\begin{equation}
1=
\frac{K_c+ \delta K}{2\gamma} \left[ \frac{\pi}{2} -\delta \theta
+\frac{1}{2}\sin(\pi-2\delta\theta)\right]
\end{equation}
An expansion of the sine function up to the cubic term yields:
\begin{equation}
0=\frac{\pi}{2}\delta K - \frac{8\gamma}{3 \pi} \delta\theta^3
\label{dkdt}
\end{equation}
Thus the problem reduces to 
finding $\delta \theta$, 
from Eq.~(\ref{mean_field}):
\begin{eqnarray}
\gamma&=&(K_c+\delta K) (r_c + \delta r) \sin(\pi/2 -\delta \theta) \\
&\Rightarrow& \delta \theta \approx 
 \sqrt{\frac{8}{\pi}\delta r + \frac{\pi}{2\gamma}\delta K}
\label{dteta}
\end{eqnarray}
Introducing this 
expression into Eq.~(\ref{dkdt}), 
a formula for the asymptotic dependence of 
$\delta r$ on $\delta K$ is obtained:
\begin{equation}
\delta r = \left[\frac{9\pi^7}{2^{17} \gamma^2}\right]^{1/3}\delta K^{2/3} -
\frac{\pi^2}{16\gamma}\delta K 
\label{2/3}
\end{equation}
The result is compared in
Fig.~\ref{scaling} to the exact solution, arising
from numerically solving Eq.~(\ref{solvnum}). It confirms that the order parameter
grows from $r_c$ with a power of $K-K_c$ with exponent 2/3. 
To our knowledge, this exponent and
expression (\ref{2/3}) 
have not been reported before.

\section{Finite population}
\label{discr}

Finite-size effects in the Kuramoto model have been previously 
considered in the literature. Among the different 
approaches, we may list: the  investigation of the divergence 
of fluctuations around criticality~\cite{daido87,daido90},
the observation of ephemeral coherent structures
in the incoherent state~\cite{balmforth2000}, 
and the  reduction to a normal form in
the case of identical natural frequencies with 
additive noise~\cite{pikarufo}. Also, very recently,
Mirollo and Strogatz~\cite{mirollo2005} have analyzed 
the (local) stability of the fully locked state for a finite
population, finding that the locked solution is 
stable and disappears
in a saddle-node bifurcation~\footnote{In the presence of 
an irrelevant degenerate eigenvalue at zero due to global
phase-shift invariance.} 
(as already observed in~\cite{maistrenko} for small $N$).

In this section, we show that for
the uniform frequency distribution finite-size effects 
---on the order parameter and on the loss of the synchronization---
can be studied in a novel way.

\subsection{Dependence of the order parameter on $K$}

For a finite population in the synchronized state ($K\ge K_s$), 
the order parameter is expressed [in correspondence with the integral
form (\ref{rintegral})] by:
\begin{equation}
r=\frac{1}{N} \sum_{j=1}^N \sqrt{1-\frac{\omega_j^2}{K^2r^2}}
\label{riemann}
\end{equation}
where the order parameter $r$
is a time independent quantity. 
We devote the following 
lines to deducing an $N$-dependent self-consistency equation 
[that reduces to (\ref{solvnum}) in the $N\rightarrow\infty$ limit]. 
It is accurate provided
that $K$ is not too close to $K_s$.    

\begin{figure}
\includegraphics[width=3.1in]{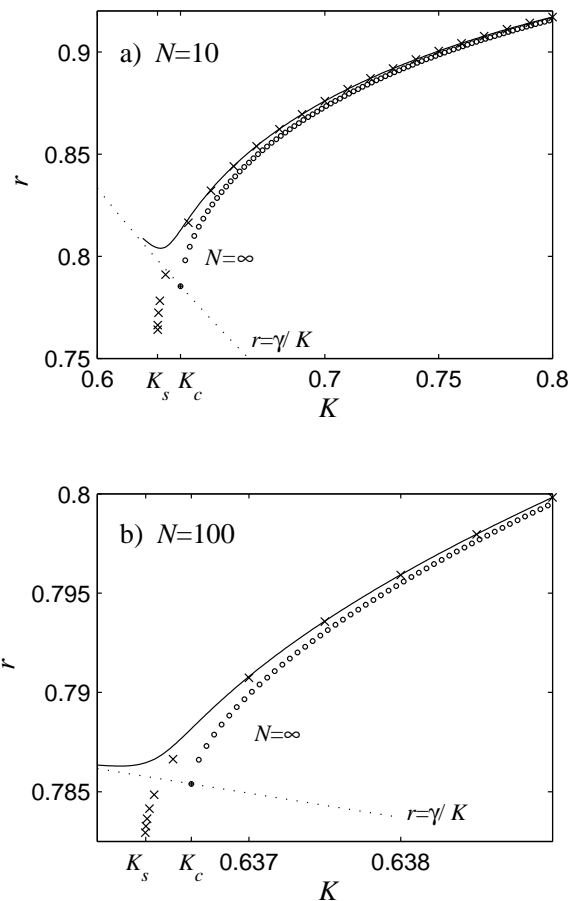}
\caption[]{Dependence of $r$ on $K$ for finite populations:  
(a) $N=10$, and (b) $N=100$; $\gamma=1/2$ in both cases. Values obtained from
a direct computation of the Kuramoto model are marked by $\times$.
The first frequency splitting is observed at $K_s$
(for $K<K_s$, $r$ is no longer constant in time). 
The solid line is obtained solving 
Eq.~(\ref{solvnumn}) numerically. It matches the 
computed values, but fails when approaching the line $r= \gamma /K$. 
As a reference, the solution for an infinite population is shown with 
circles [after numerically solving Eq.~(\ref{solvnum})]; 
the critical point $(K_c,r_c)$ is marked
with a $\bullet$ symbol.} \label{selfcheck}
\end{figure}

According to (\ref{unif}), the natural frequencies of the 
finite population are taken with step $\Delta \omega =\omega_{j+1}-\omega_j=2\gamma/N$. 
Therefore, the order parameter (\ref{riemann})
is equivalent to a Riemann sum with 
constant step of the integral found
in the continuum limit (\ref{rintegral}). The discrepancy
between both cases is, at the leading order,
dependent on $f''$  (where $f(\omega) \equiv \sqrt{1-(\omega/Kr)^2}$). 
Hence, for one Riemann box centered at $\omega_j$: 
\begin{equation}
\int_{\Delta\omega}f(\omega) d\omega = f(\omega_j) \Delta\omega +
\frac{f^{''}(\omega_j)}{24}  \Delta \omega^3 + {\cal O}(\Delta\omega^5) .
\label{serie1}
\end{equation}
Therefore, for the finite $N$ case we may approximate (\ref{riemann}) by:
\begin{equation}
r \approx  \frac{1}{2\gamma}\int_{-\gamma}^\gamma \sqrt{1-\frac{\omega^2}{K^2r^2}} d\omega
-\frac{\gamma^2}{6N^3}\sum_{j=1}^N f''(\omega_j)
\end{equation}
In this expression, the sum may be approximated
by its corresponding integral 
\begin{eqnarray}
\sum_{j=1}^N f''(\omega_j)&=&
\frac{N}{2\gamma}\left[ \int_{-\gamma}^\gamma f''(\omega) d\omega
+ {\cal O}(N^{-2}) \right] \nonumber \\ &\approx& \frac{N}{2\gamma} 
\left. f'(\omega) \right|_{-\gamma}^\gamma
=-\frac{N}{\gamma Kr\sqrt{\frac{K^2r^2}{\gamma^2}-1}}
\label{serie2}
\end{eqnarray}
And we obtain a self-consistency equation for a
finite population of $N$ oscillators:
\begin{equation}
r= \frac{1}{2} \sqrt{1-\frac{\gamma^2}{K^2r^2}}+
\frac{Kr}{2\gamma}\arcsin\left(\frac{\gamma}{Kr}\right)
+ \frac{\gamma N^{-2}}{6Kr\sqrt{\frac{K^2r^2}{\gamma^2}-1}}
\label{solvnumn}
\end{equation}
With respect to  the equation for the thermodynamic 
limit [Eq.~(\ref{solvnum})], there is an additional term 
depending on $N$. 
Plots of the numerical solutions for $N$ = 10 and 100 
are shown in Fig.~\ref{selfcheck}. 
Eq.~(\ref{solvnumn})
is deduced from the continuum equation, and therefore 
cannot intersect the line $r=\gamma / K$, because several
terms explode. Note that adding more terms of the series
in Eqs.~(\ref{serie1},\ref{serie2}) does not 
overcome this problem. This suggests that, unfortunately, 
the behavior very close to $K_s$ cannot be deduced
by simply manipulating the equations for an infinite population.
Nonetheless, out of that region,
the solution of (\ref{solvnumn}) reproduces
the numerical results, even for such a relatively
small number of oscillators as $N=10$, Fig.~\ref{selfcheck}(a).

\subsection{Thermodynamic limit of the first frequency-splitting}

\begin{figure}
\includegraphics[width=3.1in]{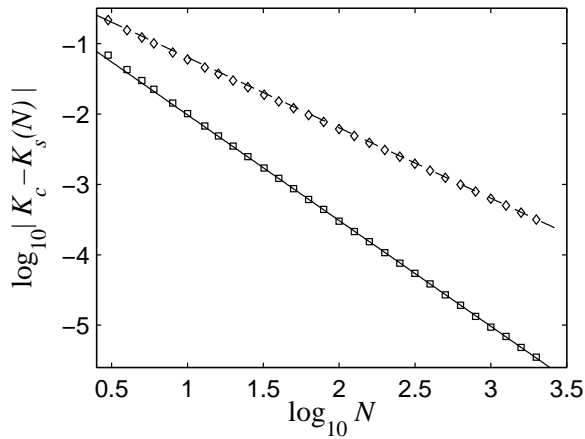}
\caption[]{Log-log dependence of the distance from $K_s$ 
to $K_c$ on the population
size ($\gamma=1/2$). Squares and diamonds correspond
to  different arrangements of the natural frequencies, 
Eqs.~(\ref{unif}) and (\ref{unif2}), respectively.
In the first case, the last two decades were 
fitted with a straight line, yielding a slope $-\mu=-1.502$.
The dashed straight line $\log_{10}(4\gamma/\pi)-x$ 
arises from Eq.~(\ref{-1}).} \label{asymp}
\end{figure}

The 
arrangement of the natural frequencies 
in Eq.~(\ref{unif}) converges to
the uniform frequency distribution (\ref{g}). 
But, as explained above, this limit is non-trivial:
a first-order phase transition is substituted, when $N$ becomes
finite,
by a set of frequency-splitting bifurcations
accumulating at $K_c$. Numerically, the study of these
bifurcations is quite involved. Nonetheless, the
point where, as $K$ decreases, 
the first splitting occurs ($K=K_s$), can be accurately computed.
This is possible because above $K_s$ the system is in a fixed point state.
Our simulations, Fig.~\ref{asymp}, indicate
that for the arrangement in (\ref{unif}) $K_s$ converges to ${K_c}^-$ according to a power law: 
\begin{equation}
K_c-K_s(N) \propto N^{-\mu} .
\label{powerlaw}
\end{equation} 
Note that $K_s$ and $K_c$ are both
proportional to $\gamma$, so $\mu \approx 1.5$ is independent of $\gamma$.

The recipe followed to mimic the thermodynamic limit 
was to divide the frequency distribution $g(\omega)$ 
in $N$ parts of equal area taking each $\omega_j$ 
at the center of 
each block. 

Nonetheless, there are many (infinite in fact) possible 
discrete arrangements with the same continuum limit, 
 but for arrangement (\ref{unif}) $\mu$ is large
enough to deduce the decay for other distributions, 
by just considering the variation of the effective $\gamma$.
For instance, if the natural frequencies
are taken as in \cite{matthews91,inertiad}:
\begin{equation}
\omega_j=-\gamma+\frac{\gamma}{N-1}2(j-1)
\label{unif2}
\end{equation}
one observes that $K_s$ converges
to ${K_c}^+$  with a power-law, see $\diamond$'s in Fig.~\ref{asymp}, 
but more slowly than 
for the arrangement in (\ref{unif}).
As the extrema of (\ref{unif2}) are fixed,
a change in $N$ varies the effective width of 
the equivalent continuous distribution: 
$\gamma_{eff}=\gamma + \gamma/(N-1)$.
Hence, from (\ref{powerlaw}) we get:
\begin{equation}
K_s(N)=\frac{4\gamma_{eff}}{\pi}-\alpha N^{-\mu}
\approx K_c + \frac{4 \gamma}{\pi N} +{\cal O}(N^{-\mu})
\label{-1}
\end{equation}
which agrees with the observed result 
(see the dashed line in Fig.~\ref{asymp}).

\section{Other frequency distributions}
\label{other}

In this section, we briefly discuss the extension of the previous 
results to other 
frequency distributions supported on a finite interval $[-\gamma,\gamma]$.
We first note that for non-uniform 
distributions the lost of complete synchronization and 
the critical point where the incoherent solution becomes unstable 
do not coincide: 
$K_s^\infty \equiv K_s (N\rightarrow \infty) > K_c$. In other words,
there exists an intermediate range of partial entrainment in which
one part of the population is synchronized whereas the remaining oscillators
drift.

As model distributions we considered 
three  unimodal distributions listed in Table~\ref{tabla}.
As for the uniform distribution, the desynchronization point $K_s^\infty$ 
may be computed analytically using Eq.~(\ref{rintegral}).  

\begin{table}
\centering
\begin{tabular}{l  c  c} \hline\hline
 & $g(\omega)$ & $K_s^\infty$  \\ \hline
Parabolic &  $\frac{3}{4 \gamma^3}(\gamma^2-\omega^2)$  & $\frac{32\gamma}{9 \pi}$ \\ \hline
Triangular & $\frac{\gamma - |\omega|}{\gamma^2}$ & $\frac{6 \gamma}{3\pi - 4}$ \\ \hline 
Hat-shaped & 
$ \begin{array}{l} \frac{2}{3\gamma} \quad (|\omega| \leq \frac{\gamma}{2})\\ 
\frac{1}{3\gamma} \, (\frac{\gamma}{2} < |\omega| \le \gamma) \end{array}$&  
$\frac{36 \gamma}{ 8\pi +3\sqrt{3}}$ \\ \hline\hline 
\end{tabular}
\caption{Three frequency distributions considered in 
Sec.~\ref{other}.} 
\label{tabla}
\end{table}

We focused on two simple sampling schemes
to discretize distributions supported on a bounded interval:
\begin{enumerate}

\item[(i)] $\int_{\tilde\omega_j}^{\tilde\omega_{j+1}} g(\omega) d\omega= 2\gamma /N ,\tilde\omega_1=-\gamma, 
\omega_j=\frac{\tilde\omega_j+\tilde\omega_{j+1}}{2}$.
\item[(ii)] $\int_{\omega_j}^{\omega_{j+1}} g(\omega) d\omega= 2\gamma /(N-1) , \omega_N=-\omega_1=\gamma$.
\end{enumerate} 
Applying (i) and (ii) to a uniform distribution one gets the 
arrangements in Eqs.~(\ref{unif}) and (\ref{unif2}), respectively.

For both schemes and the three frequency distributions
in Table~\ref{tabla}, 
the approach of the first frequency splitting
to the thermodynamic limit satisfies 
a power law: $|K_s^\infty-K_s(N)|\propto N^{-\mu}$,
as occurred for the uniform distribution 
($K_s^\infty=K_c$ in this case).
For sampling (ii) we find that the value of the exponent is
always $\mu \approx 1$. However for (i), different exponents
arise, in contrast to $\mu \approx 1.5$ obtained for the uniform distribution:
$\mu\approx 0.5$ for triangular and parabolic distributions,
and $\mu\approx 1$ for the hat-shaped one. We have checked that this exponent
arises for other 
distributions with an abrupt boundary ($g(\pm \gamma)>0$). 

Another interesting power law is the shift
of the order parameter in the synchronized state:
$|r(K,N)-r(K,N=\infty)| \sim  N^{-\nu}$.
For sampling scheme (i) 
theoretical results may be obtained, using again arguments based
on the Riemann sum.
If $g(\pm \gamma) > 0$  ---e.g.~uniform [see Eq.~(\ref{solvnumn})] 
or hat-shaped distributions--- one may obtain 
a formal solution \footnote{$r= \int_{-\gamma}^\gamma g \, f \,  d\omega 
- \frac{N^{-2}}{24} \int_{-\gamma}^\gamma \frac{g  f'' \:   
+ 2 g' f'}{g^2} d\omega$.} that yields
$\nu =2$. 
However for distributions that approach zero 
at $\omega=\pm\gamma$, the ``Riemann-sum approach'' is not
valid due to divergences at $\omega=\pm \gamma$. One must,
therefore, analyze these points separately. In particular,
one obtains $\nu=3/2$ for linearly decaying
distributions [$g(\omega \rightarrow \pm \gamma) \sim (\gamma \mp \omega)$, 
e.g. parabolic and triangular distributions].  Also, our simulations
indicate that for sampling (ii) $\nu \approx 1$, irrespective of the 
frequency distribution (for the uniform distribution $\nu=1$ is straightforward
due to the ${\cal O}(N^{-1})$ effective shift of $\gamma$).

Finally, we note that a recipe similar to (i) consisting in
taking
the frequencies at the median (instead of the center) of each 
block~\footnote{$\int_{\omega_j}^{\omega_{j+1}} g(\omega) d\omega= 2\gamma /N= 
2\int_{-\gamma}^{\omega_1} g(\omega) d\omega$.}
exhibits the same exponents $\mu, \nu$ that scheme (i).

\section{Conclusions}
\label{concl}
In the present paper, the first-order phase transition arising
when imposing a uniform frequency distribution on  
the Kuramoto model has been studied.
In the case of an infinite population, 
we have found an explicit asymptotic 
dependence of the order parameter after criticality, Eq.~(\ref{2/3}).

For a {\em finite} population, our first conclusion
is that in contrast to strictly unimodal distributions
(Cauchy, Gaussian, parabolic,...) of the natural frequencies, 
which exhibit transitions
of second-order type,
the thermodynamic limit is non-trivial for a uniform 
distribution. In the finite-$N$ case, 
the synchronized state does not split directly into $N$ clusters,
but through a cascade of frequency splittings. To be 
congruent with the first-order phase transition
predicted in the thermodynamic limit all the splittings
must accumulate at $K_c$ as $N\rightarrow\infty$.

The dependence of the order parameter $r$ 
on the coupling $K$, has been 
expressed in an easy-to-compute formula, Eq.~(\ref{solvnumn}).
In this formula, the population size $N$ enters explicitly,
and it allows, except very close to the desynchronization
point, 
an accurate computation of the order parameter,
even for small $N$.  

{Two sampling schemes to set the natural
frequencies have been compared. 
The scheme we propose in (\ref{unif}) converges to the 
thermodynamic limit faster than another used in the literature [Eq.~(\ref{unif2})]. 
The comparison is
based on the different exponent of the power-law convergence 
for the point of the first frequency splitting ($\mu\approx 1.5$ vs. $\mu =1$) and the 
shift of the order parameter ($\nu=2$ vs. $\nu=1$). 
Other
frequency distributions (with compact support) different from the uniform one
have been discussed. Among the infinite
possible sampling schemes, those studied here appear the most 
natural ones to us. Nonetheless, further investigation is needed 
to assess the existence of an optimal
sampling scheme to mimic the thermodynamic limit
with a finite population.}



In spite of the lack of structural stability under perturbations 
of the uniform frequency distribution (in
the thermodynamic limit, but not in the finite case 
as proved in~\cite{maistrenko}), the results in this paper 
could be useful in order to
understand more complicated
schemes, like oscillator networks~\cite{Watts}. The 
use of a frequency distribution with an
abrupt transition seems more suited to
better resolve critical points
in this kind of systems.

\acknowledgments

The author thanks Ernest~Montbri\'o for fruitful discussions and critical 
reading of the manuscript. Useful comments from Eduardo G.~Altmann 
and Luis G.~Morelli are also gratefully acknowledged.

\bibliographystyle{prsty}


\end{document}